\documentclass[letterpaper,twocolumn,10pt]{article}
\usepackage{usenix}
\usepackage{mathptmx} 
\usepackage{subcaption}
\usepackage{graphicx}
\usepackage{enumitem}
\usepackage{comment}
\usepackage{tikz}
\usepackage{amsmath}
\usepackage{threeparttable}
\usepackage{filecontents}
\usepackage{fancyhdr}
\usepackage[normalem]{ulem}

\begin{document}
	
	\date{}

\title{DP-DNA: A Digital Pattern-Aware DNA Storage System to Improve Encoding Density} 
\author{Bingzhe Li$^{\dagger}$, Li Ou$^{*}$, and David Du$^{*}$\\
	$^{\dagger}$Oklahoma State University\\
	$^{*}$University of Minnesota, Twin Cities\\
	bingzhe.li@okstate.edu; 	\{ouxxx045, du\}@umn.edu}

\maketitle


\begin{abstract}

With the rapid increase of available digital data, we are searching for a storage media with high density and capability of long-term preservation. Deoxyribonucleic Acid (DNA) storage is identified as such a promising candidate, especially for archival storage systems. However, the encoding density (i.e., how many binary bits can be encoded into one nucleotide) and error handling are two major factors intertwined in DNA storage. Considering encoding density, theoretically, one nucleotide (i.e., A, T, G, or C) can encode two binary bits (upper bound). However, due to biochemical constraints and other necessary information associated with payload, currently the encoding densities of various DNA storage systems are much less than this upper bound. Additionally, all existing studies of DNA encoding schemes are based on static analysis and really lack the awareness of dynamically changed digital patterns. Therefore, the gap between the static encoding and dynamic binary patterns prevents achieving a higher encoding density for DNA storage systems.


In this paper, we propose a new Digital Pattern-Aware DNA storage system, called DP-DNA, which can efficiently store digital data in the DNA storage with high encoding density. DP-DNA maintains a set of encoding codes and uses a digital pattern-aware code (DPAC) to analyze the patterns of a binary sequence for a DNA strand and selects an appropriate code for encoding the binary sequence to achieve a high encoding density. An additional encoding field is added to the DNA encoding format, which can distinguish the encoding scheme used for those DNA strands, and thus we can decode DNA data back to its original digital data. Moreover, to further improve the encoding density, a variable-length scheme is proposed to increase the feasibility of the code scheme with a high encoding density. Finally, the experimental results indicate that the proposed DP-DNA achieves up to 103.5\% higher encoding densities than prior work.
\end{abstract}

\section{Introduction}

With the rapid increase of available digital data, the demand for storage capacity is significantly increased. Based on the prediction of International Data Corporation (IDC)~\cite{reinsel2018digitization}, the total amount of digital data in the whole world will reach 175 Zettabyte (ZB) in 2025. To satisfy such a demand, many emerging storage technologies and devices have been exploited and developed, such as magnetic tape~\cite{tape}, Solid-State Drive (SSD)~\cite{lee2002effects, li2019haml}, Shingled Magnetic Recording (SMR)~\cite{amer2011data, wu2019zonealloy} and Interlaced Magnetic Recording (IMR) drives~\cite{hajkazemi2019track, wu2018data} in past decades. However, these storage technologies still cannot satisfy the increased demand for storage capacity. Besides, these storage devices cannot preserve data for a long duration (beyond 15 years).

Deoxyribonucleic Acid (DNA) storage is one promising candidate for archiving both a high storage density and a long preservation duration. As indicated in~\cite{appuswamy2019oligoarchive}, DNA storage can achieve a theoretical density of about 455 Exabytes/gram and DNA can be preserved for several centuries~\cite{allentoft2012half, grass2015robust}. Thus, DNA storage provides a great potential to store a massive amount of data and achieves a much higher storage density and a longer preservation duration than any other existing storage technologies. 

In the past, researchers investigated and demonstrated the feasibility of DNA storage~\cite{luby2002lt, mackay2005fountain, wang2019high, gibson2010creation, grass2015robust, heckel2017fundamental, organick2018random, church2012next, blawat2016forward, bornholt2016dna, choi2018addition}. They proposed different encoding schemes (converting binary digital data to DNA sequences), error-correction codes, and biochemical techniques for improving the efficiency of synthesis (write data to DNA) and sequencing (read data out from DNA) from both biology and information theory perspectives. However, these existing studies did not thoroughly investigate the potentials from the following two aspects. The first one is that all previous DNA storage architectures only use one encoding scheme in their systems, and thus their encoding densities are limited by the properties of the single encoding scheme. The second one is that existing encoding schemes rarely considered the data patterns in digital data. The gap between their static encoding and dynamic binary patterns makes them miss an opportunity of further improving the encoding density of DNA storage. Thus, it is important and has a great potential to build the bridge between dynamic digital data and DNA encoding to improve encoding density of DNA storage.


In this paper, we propose a \underline{D}igital \underline{P}attern-aware DNA storage system, called DP-DNA. The primary goal of the proposed DP-DNA scheme is to improve the DNA storage density by using a multi-encoding scheme to adapt to the incoming digital sequences. First, a Digital Pattern-Aware Code (DPAC) is proposed to analyze a binary sequence and distinguishe the high-frequency and low-frequency binary patterns. Then, the DPAC scheme selects an appropriate code to encode the binary sequence based on the outcome of pattern frequency analysis. To further improve the encoding density of DNA storage, DP-DNA applies a variable-length (VL) scheme to dynamically find a proper DNA strand length for a higher encoding density. As a result, the DP-DNA can achieve a high storage density according to the digital pattern of each binary data sequence for a DNA strand. To avoid certain biochemical constraints in DNA storage, a feasibility checker is used to ensure free-violations with DNA's biochemical constraints. 


The contributions of this paper are fourfold: \textbf{I.} We propose a new DNA storage architecture with using multiple encoding schemes for one DNA storage system to improve the encoding density. \textbf{II.} The pattern-aware scheme analyzes the considered digital binary sequence for a DNA strand and selects a proper code for the sequence to further improve the overall encoding density based on the distribution of binary sequence's patterns. \textbf{III.} A variable-length scheme is used to further increase the DNA storage density. \textbf{IV.} Different applications including images, database, etc. are used to test the encoding density of DNA storage, and the final results indicate the proposed scheme achieves much higher encoding densities than prior work.

The remainder of this paper is organized as follows: Section~\ref{sec:background} provides background information about DNA storage. Section~\ref{sec:motivation} demonstrates the motivation behind this work. Section~\ref{sec:design} describes the proposed DP-DNA scheme. The discussion of overhead, feasibility and random access is discussed in Section~\ref{sec:overhead}. Section~\ref{sec:result} shows the comparisons of the experimental results of DP-DNA and some existing schemes. The related work is introduced in Section~\ref{sec:related}. The conclusions are drawn in Section~\ref{sec:conclusion}.

\section{Background of DNA Storage}\label{sec:background}
\begin{figure}[!t]
	\centering
	\includegraphics[width=3.3in]{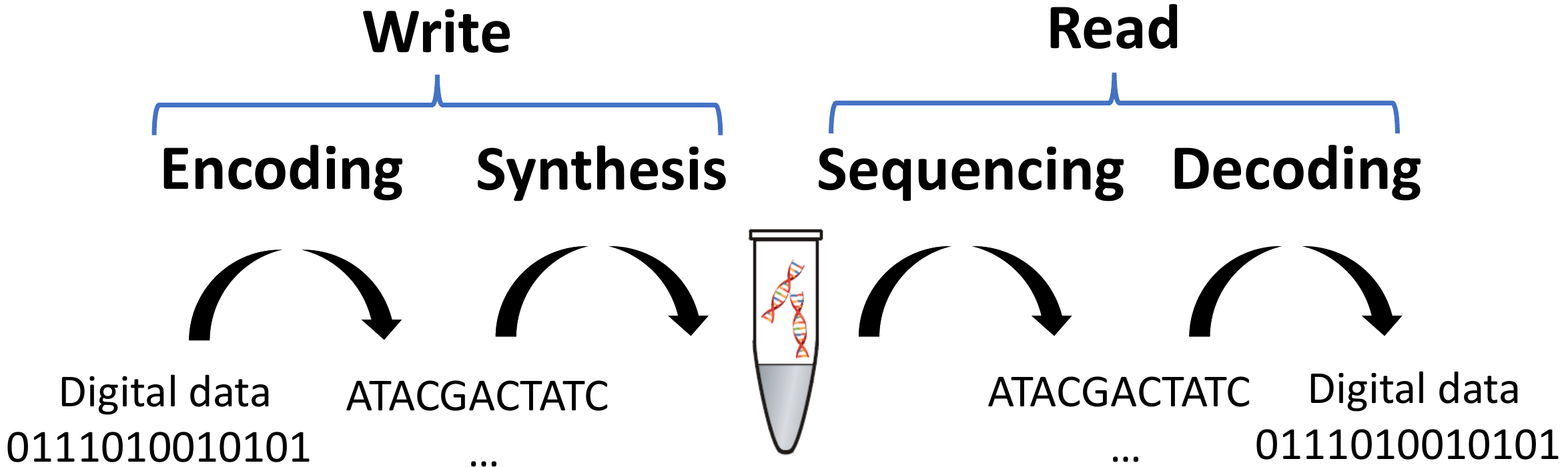}
	\caption{Basic steps of DNA storage.}
	\label{fig:basic_step}
\end{figure}

In DNA, nucleotides are the basic molecules to build DNA sequences. Each DNA nucleotide contains one of four bases: Adenine (A), Cytosine (C), Guanine (G), or Thymine (T). A DNA strand or oligonucleotide consists of a chain of these different nucleotides. Two complementary DNA strands bind together to form a DNA molecule in the famous double helix structure. An A in one strand will only bind with a T in the other, and vice versa. Similarly, a C will only bind with a G, and a G will only bind with a C. The number of these base pairs (bp) in the DNA molecule is a measurement of the DNA sequence's length. A DNA sequence is a list of the four base letters in the order in which they appear in a DNA strand. For our purposes, saying nucleotides refers to these four letters. As shown in Figure~\ref{fig:basic_step}, there are four major processes in the DNA storage system: encoding, synthesis, sequencing, and decoding. The encoding and synthesis are the processes to write data into DNA storage. The sequencing and decoding are the processes to read data out from DNA storage.

{\bf Encoding and decoding:} To store data in DNA storage, we need to convert binary digital data into DNA sequences (i.e., A, T, G, and C). Ideally, we can store two binary bits in each base pair (e.g., 10->A, 11->T, 01->C, and 00->G). However, due to some biochemical constraints such as homopolymers (repetitions of the identical nucleotide like AAAA...), the two-bit encoding scheme introduces a much higher error rate in both synthesis and sequencing processes. Therefore, most existing studies~\cite{church2012next, goldman2013towards, grass2015robust, bornholt2016dna, erlich2017dna, blawat2016forward, organick2018random, anavy2018improved, choi2018addition, lee2018enzymatic} convert digital data to DNA sequences with less than 2 bits/nt (bits per nucleotide). Decoding is the reverse of the encoding process, which transfers the DNA sequence back to original binary data according to the used encoding scheme.

{\bf DNA synthesis (write) and sequencing (read): } After converting binary data to DNA sequences, we can chemically synthesize a DNA sequence nucleotide by nucleotide as the designs in~\cite{kosuri2014large, matteucci1981synthesis}. The process is error-prone since a nucleotide can be aligned to an existing partial strand. Current technologies can synthesize a DNA strand up to 3,000 base pairs (bp)~\cite{yazdi2015rewritable,HosseinTabatabaeiYazdi2017,GeneArt}. After synthesis, millions of different DNA strands representing binary data are mixed in one tube/pool. The target DNA strands are amplified/duplicated during the read operation via Polymerase Chain Reaction (PCR). PCR is a biotechnological method for exponentially duplicating target DNA strands within a tube. After that, the sample of DNA strands is sent to a sequencing machine. The target DNA strands as a template will be duplicated by attaching fluorescent nucleotides with different colors. Finally, the target DNA strands will be read out by the sequencing machine. Please note that multiple different DNA strands can be read out in a massively parallel way if they are encoded with the same primer pair (primer pairs will be explained in the next section). In a PCR process, all strands encoded with the same primer pair are the target DNA strands.

{\bf DNA Storage Errors:} In DNA storage, there are three major types of errors including deletion, insertion and substitution. More specifically, some nucleotides might be written or read by another type of nucleotides (i.e., substitution) or may not be written into the DNA strands or read out (i.e., deletion) or newly added into or read out (i.e., insertion). The substitution error rate may reach 0.008 errors per base, which are around 2X-10X higher than the other two types~\cite{heckel2019characterization, organick2018random}. Moreover, some features may increase the error probability in synthesis and sequencing processes. For example, due to the technology limitation, with increasing DNA strand length, it becomes harder to add more nucleotides on the DNA strands. In other words, when the strand length increases, the errors happening on each nucleotide bind also exponentially increase~\cite{Richterich1998, GeneArt, yazdi2015rewritable, organick2018random, erlich2017dna, li2020can}. Therefore, most of the existing works for DNA storage use 100\textasciitilde{}300 bp length of a DNA strand. GC content (i.e., the percentage of bases in a DNA sequence that are either C or G) is related to the melting temperature of DNA strands during PCR. Too high and too low GC content cause too high melting point and unspecific binding with primers, respectively, which makes PCR more difficult and error-prone. 

Some other features such as long homopolymers (e.g., AAAAA), hairpins/loops and other forms of higher-order structure~\cite{nelms2011predicted} also induce the difficulty of writing or reading DNA strands. Although hairpins/loops and higher-order structure increase the difficulty of sequencing, DNA strands with those structures are possibly read out by one or several time sequencings. Therefore, most of the existing DNA storage studies~\cite{church2012next, goldman2013towards, grass2015robust, bornholt2016dna, erlich2017dna, blawat2016forward, organick2018random, anavy2018improved, choi2018addition, lee2018enzymatic, chandak2019improved} proposed their encoding schemes only intentionally avoiding the long homopolymers and low/high GC content since the hairpins/loops and other forms of a higher-order structure are hard to avoid and might be mitigated by sequencing multiple times.

\begin{figure}[!t]
	\centering
	\includegraphics[width=3.3in]{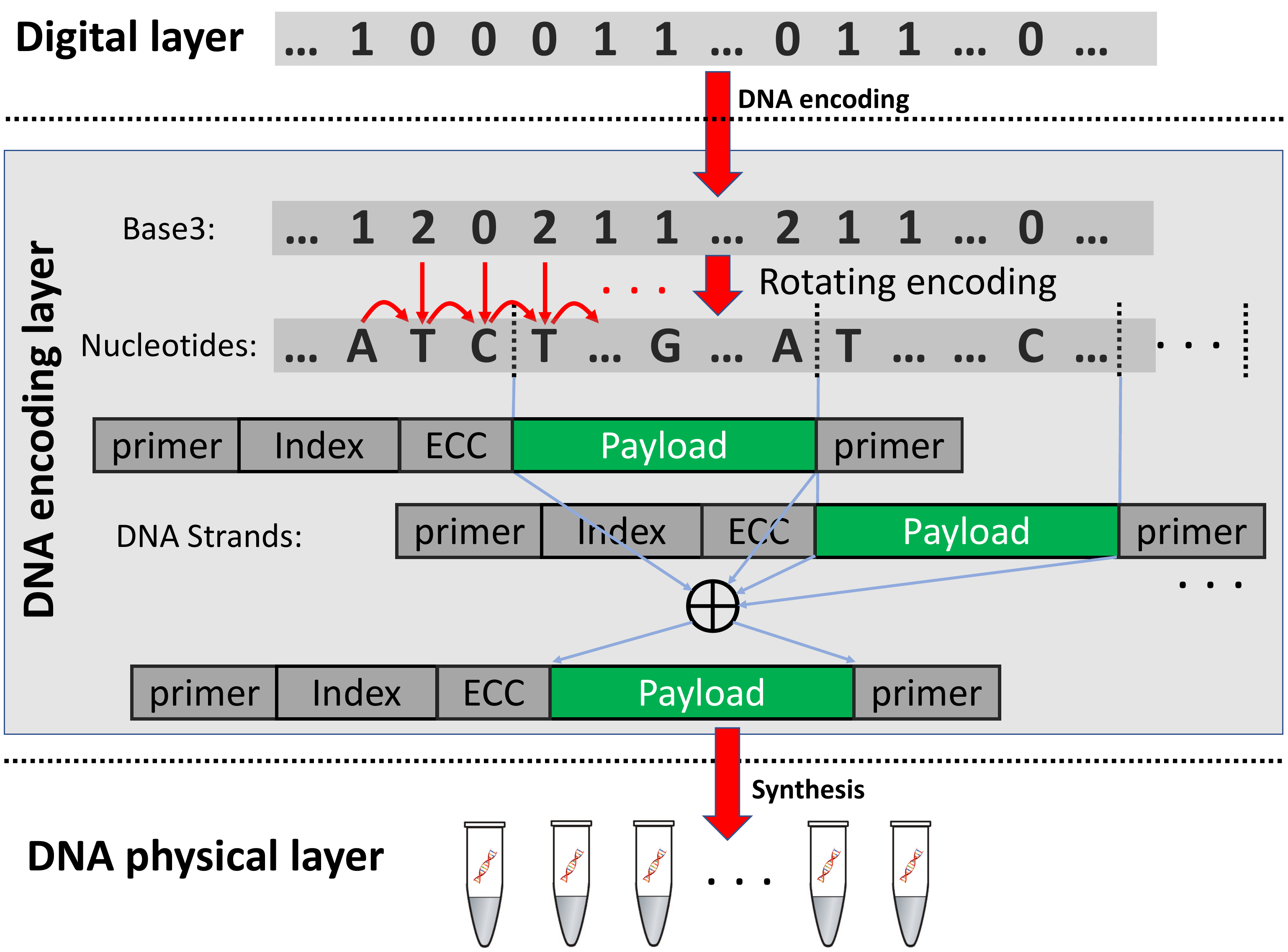}
	\caption{An example of DNA encoding process~\cite{bornholt2016dna}.}
	\label{fig:architecture}
\end{figure}

{\bf Overall encoding process of DNA storage and encoding format: }Figure~\ref{fig:architecture} provides an example of the whole DNA encoding process to encode binary data to DNA sequences~\cite{bornholt2016dna}. First, the binary data is encoded to base-3 codes. This process is to compress data to achieve a higher encoding density. A rotating code is used to encode digital data into nucleotides based on its immediate previously encoded nucleotide and current digital values. This rotating process is to avoid homopolymers and balance the GC contents. After that, the long DNA sequence will be chunked into short DNA sequences (e.g., 50 bp - 250 bp), which are the payloads of DNA strands. Then, some necessary DNA segments such as primers and an index are attached to a payload to comprise a DNA strand.  

In Figure~\ref{fig:architecture}, a DNA strand follows the format including three fields: primer, index, error correction code (ECC), and payload. A primer is a short nucleotide sequence with a length from 18\textasciitilde{}25 bp~\cite{primerdesign}. One primer is normally attached to the beginning and another to the end of a DNA strand. Therefore, we refer these two primers as a primer pair. One primer pair can be associated with thousands or millions of different DNA strands since they can be read out in one sequencing. To distinguish these DNA sequences, an internal index field is added in the DNA strand. ECC (error-correction code) is used to correct errors and recover original data since both synthesis and sequencing are error-prone. Some schemes~\cite{bornholt2016dna} apply ECC across different DNA strands. As indicated in wet-lab experiments~\cite{organick2018random}, to make DNA sequences to be correctly recovered, and make fair comparisons, we apply a default 15\% ECC overhead for all schemes in this paper (for other ECC overheads, the conclusion of this paper will not change). The payload is the data information.

\section{Motivation}\label{sec:motivation}
\begin{figure}[!t]
	\centering
	\includegraphics[width=2in]{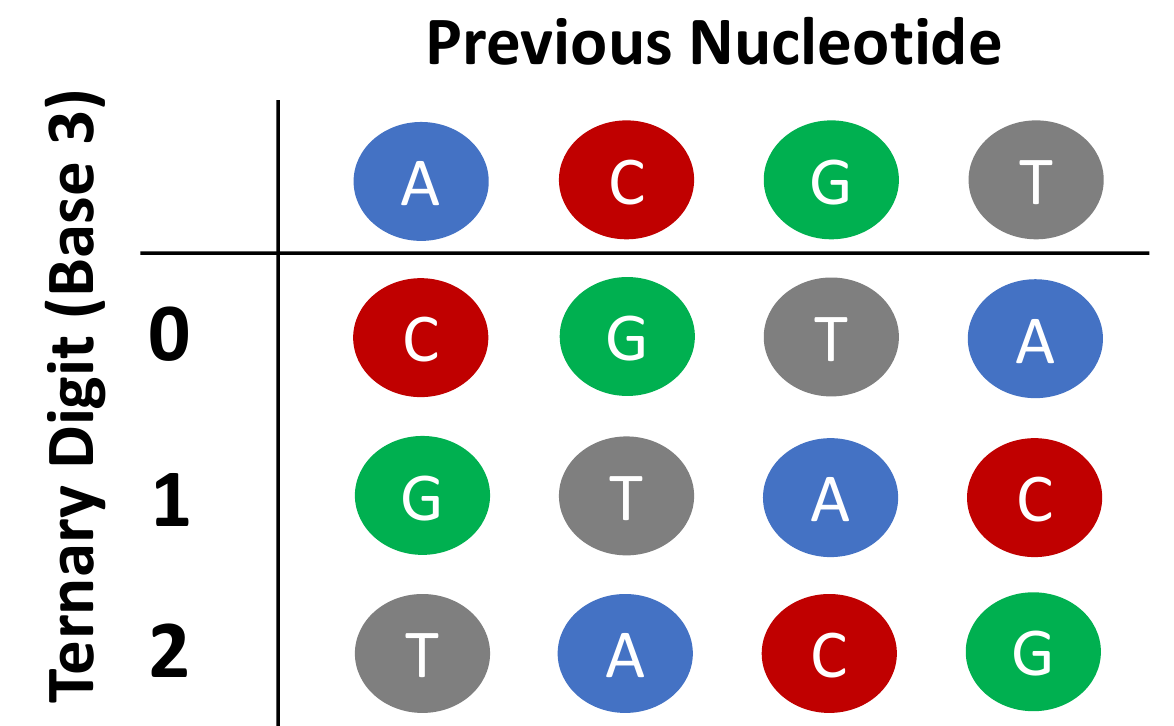}
	\caption{A rotating encoding scheme converts ternary digits (i.e., 0, 1, and 2) to nucleotides~\cite{blawat2016forward, bornholt2016dna}.}
	\label{fig:rotation_code}
\end{figure}
In this section, we introduce the motivation of our work. The encoding density is defined by the number of digital bits stored in DNA storage divided by the number of nucleotides (nt) used. Ideally, the encoding density is 2 bits/nt (bits per nucleotide). However, since other fields such as primer and internal index are also necessarily associated with payload in one DNA strand, the encoding density normally is smaller than 2 bits/nt. Even if we only consider the payload's encoding density, the DNA storage struggles to achieve 2 bits/nt due to some biochemical constraints such as the absence of homopolymers and preferring a certain GC content. Considering these biochemical constraints, an upper bound encoding density of 1.98 bits/nt has been theoretically proven~\cite{erlich2017dna}. However, the upper bound (i.e., 1.98 bits/nt) was achieved by filtering out many data (only encode up to 500MB data). In other words, the scheme has the scalability issue and cannot be used to a large DNA tube capacity such as storing GB or TB level data, which is contradicted to the purpose of DNA storage. Moreover, their work involved random numbers to shuffle the original data but introduced a large overhead of storing those metadata (i.e., random numbers) to recover binary values from DNA sequences. In real-world storage systems, a large amount of incoming sequences with dynamic access patterns prevent DNA storage from reaching this upper bound. 

For other types of DNA storage systems, they used the schemes mapping from digital data to DNA sequences based on a static view. For example, the state-of-the art encoding scheme widely used in many previous studies~\cite{goldman2013towards, organick2018random, blawat2016forward, tomek2019driving} converts each byte in the binary sequence to several trits (base-3 digits). The purpose of converting to trits values (i.e., 0, 1, and 2) is that they can apply a rotation code (i.e., mapping 0, 1 and 2 to A, T, C and G based on the previous nucleotide as shown in Figure~\ref{fig:rotation_code}) to avoid consecutive DNA patterns or homopolymers since there is no two consecutive identical nucleotides. For one byte binary data, there are 236 5-bit trits and 20 6-bit trits in their base-3 digit mapping table. The static coding mapping table is originally generated based on several specific files~\cite{goldman2013towards}. So, on average, they can achieve an encoding density of 1.57 bits/nt. However, these studies did not consider the patterns in digital binary sequences before encoding them into DNA sequences. For example, if a pattern `11' frequently appears in a binary sequence for a DNA strand, we may want to use a higher encoding density code for the pattern `11' to increase the encoding density. The code mapping used in the previous studies is based on a fixed set of bytes (i.e., from one long binary sequence or a dataset). As a result, it may not produce the best outcome for short sequences and input data with dynamically changed binary patterns. Therefore, current existing schemes based on statical encoding manners are hard to adapt to the dynamic workloads and thus either achieve low encoding densities or only can apply for small amount of data.


\section{Design of DP-DNA}\label{sec:design}
In this section, we start to introduce the design of DP-DNA and describe how DP-DNA can improve the encoding density of DNA storage based on the dynamically changed digital patterns. 

The encoding process as indicated in Figure~\ref{fig:basic_step} is to encode a binary sequence for a DNA strand to a DNA sequence (i.e., A, T, C, and G). As discussed in the previous sections, the design should follow some biochemical constraints such as the absence of homopolymers and low/high GC content (percent of the DNA sequence that is G or C). According to the studies of synthesis and sequencing errors in DNA storage~\cite{schwartz2012accurate, ananda2013distinct, erlich2017dna, DNA_rules}, the GC content and homopolymer runs are two significant determinants for causing errors. Essentially, the GC content should be between 40\% - 60\%, and homopolymer runs should be less than four nucleotides (i.e., the number of consecutive identical nucleotides should be less than four).

\subsection{Unbalanced Code}
\begin{figure}[!t]
	\centering
	\includegraphics[width=3.3in]{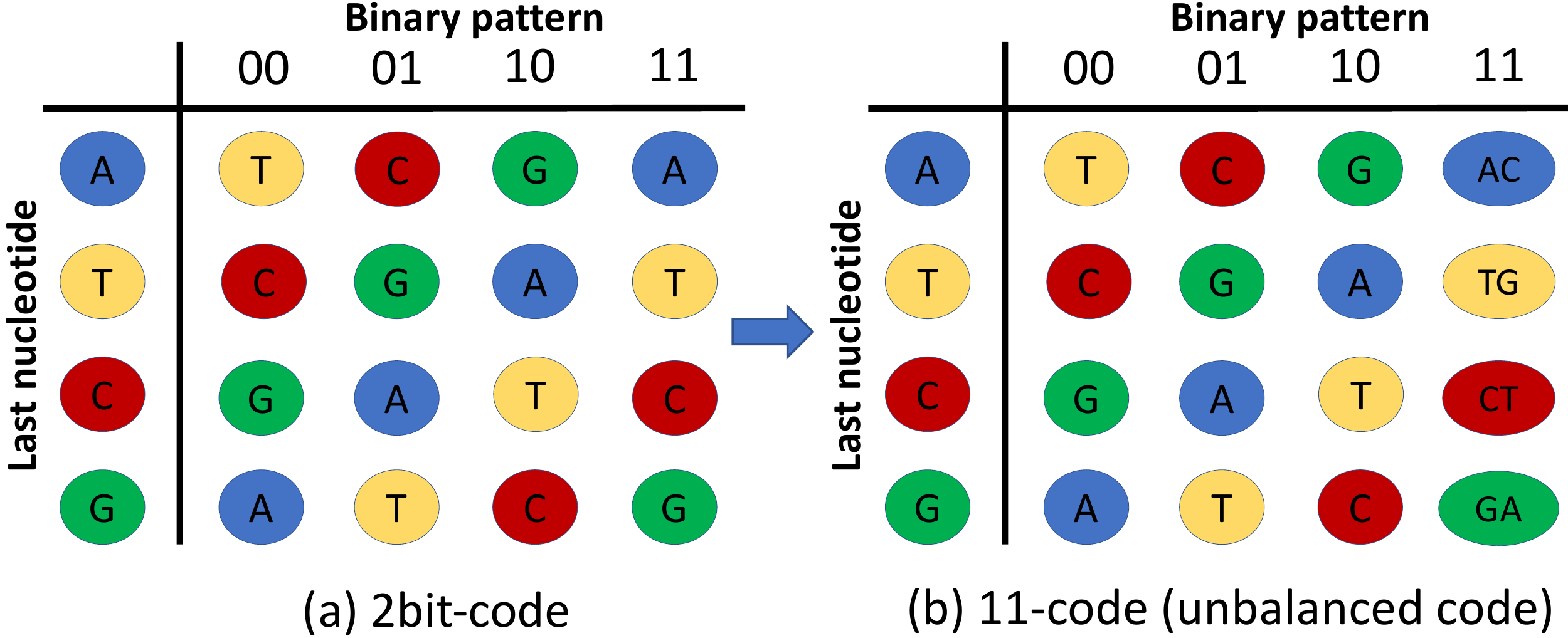}
	\caption{(a) straightforward rotating code (called 2bit-code) with the encoding density of 2 bits/nt. (b) unbalanced code (also called 11-code referring to the `11' column having two nucleotides and other columns having only one nucleotide).}
	\label{fig:coding_combined}
\end{figure}

According to the above constraints, we start with a simple proposed rotating code as shown in Figure~\ref{fig:coding_combined} (a). This code is motivated by the theoretical upper bound of encoding density and the rotating code used in~\cite{bornholt2016dna}. We try to encode two binary digits into one nucleotide and keep the nucleotide rotating to avoid long homopolymers. According to the figure, two binary digits are encoded based on their values and the previously (last) encoded nucleotide. For example, if the last DNA nucleotide is `C', the current digits`10' will be encoded into a DNA nucleotide `T'. This process continues until all binary digits finish encoding. The encoding algorithm provides two encouraging properties. One is that the encoding density obtains the upper bound of 2 bits/nt. The other one is that the probabilities of different nucleotides are the same in the encoding chart as shown in Figure~\ref{fig:coding_combined}. Thus, it can achieve around 50\% GC content, which satisfies the biochemical constraints on GC content. However, one weakness of the code is that the absence of homopolymers less than 4nt cannot be guaranteed. For example, if we have eight consecutive '1's with a former nucleotide `A', the DNA sequence output will be 'AAAA' which violates the DNA storage constraints and may induce a high error rate.

To avoid the long homopolymers, we change the encoding scheme by adding one more nucleotide into the column that may induce a long homopolymer. As shown in Figure~\ref{fig:coding_combined} (b), the last column (column '11') is mapped to two nucleotides. By doing so, it avoids the scenario of a long homopolymer. Compared to the base-3 Huffman encoding scheme~\cite{goldman2013towards}, on average, it improves the encoding density from 1.57 bits/nt to 1.6 bits/nt (a further improvement is discussed in the following sections). The small improvement in the encoding density can significantly save archiving storage capacity and cost due to the huge volume of data. Moreover, this improvement does not introduce any overhead compared to the previous work. Not only that, but it also saves the space of storing the Huffman table and encoding time compared to prior work~\cite{goldman2013towards}. 

\begin{figure}[!t]
	\centering
	\includegraphics[width=3.3in]{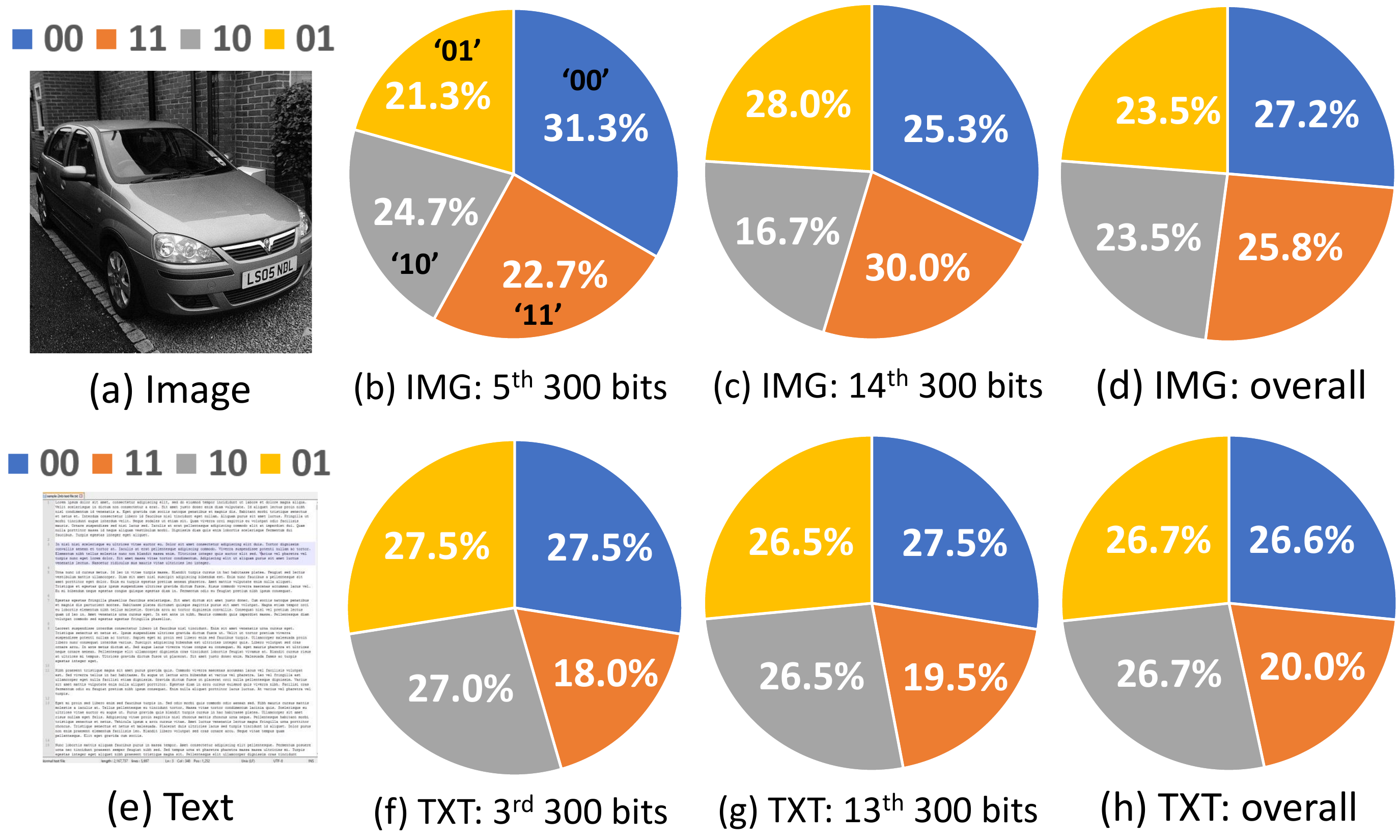}
	\caption{Two examples of the distributions of binary patterns (`11', `10', `01' and `11') in different sequences of one image~\cite{imagenet} (a - d) and one sample text file~\cite{textfile} (e - h).}
	\label{fig:digit_distribution}
\end{figure}
\subsection{Digital Pattern Aware Code (DPAC)}
Based on the unbalanced code, binary digital patterns can be converted to one nucleotide or two nucleotides based on the previous encoded nucleotide and current 2-bit digits as shown in Figure~\ref{fig:coding_combined} (b). According to the figure, 2-bit digits may be encoded to two nucleotides (i.e., 1 bit/nt for these 2-bit digits), thus decreasing the overall encoding density. One extreme case is that all 2-bit digits are encoded to two nucleotides, and it achieves the overall encoding density of 1 bit/nt. Therefore, although the unbalanced code on average achieves a little higher encoding density than the prior work, it cannot guarantee a higher encoding density than the prior work. 

Moreover, we investigate the ratios of binary data patterns (i.e., the numbers of '00', '10', '01' and '11' appeared) in a binary sequence for a DNA strand. The ratio indicates the frequency of binary patterns in a binary sequence, which is defined as the number of times a binary pattern appears in a binary sequence divided by the total number of patterns in this sequence. For example, a binary sequence `11101101' totally has four patterns (i.e., total eight bits divided by two bits per pattern equals to 4), and the pattern `11' appears twice in the sequence (i.e., $1^{st}$ plus $2^{nd}$ bits, and $5^{th}$ plus $6^{th}$ bits). So, the pattern `11' has a ratio of 50\% in this binary sequence. To investigate the ratios of the binary patterns, we find that the appearance of two consecutive binary digits mostly follows a non-uniform distribution in both short and long binary sequences. In other words, the binary patterns ('00', '10', '01' and '11') have different ratios in a sequence. Two examples in Figure~\ref{fig:digit_distribution} indicate that binary sequences have non-uniform distributions for the four digital patterns. For example, as shown in Figure~\ref{fig:digit_distribution} (c), for the $14^{th}$ 300 bits of an image, the ratio of the pattern `11' is 30\%, which is 13.3\% higher than that of the pattern `10'. So, if we follow the encoding scheme as shown in Figure~\ref{fig:coding_combined} (b), the frequently appeared pattern `11' will likely be encoded with an encoding density of 1 bit/nt. As a result, it significantly reduces the overall encoding density of DNA storage. This observation can also be applied to other applications such as text, video, etc. Based on our experiments the digit pattern distributions of different applications per 300 bits can be varied from 9\% to 75\%. Even for some data that may have overall uniform distribution, within a small fragment like 300 bits, it has high probabilities of having biased distributions for the four patterns. For example, as indicated in Table~\ref{tab:pdf}, the overall distributions of different patterns in a PDF dataset are around 25\%. However, when the data are chunked into 300-bit sequences, the pattern distributions in 300-bit sequences are varied a lot. Similar results are also found in other data types such as video, database, images, etc.

\begin{table}[!t]
\small
\centering
\tabcolsep = 3pt
\caption{The distributions of binary sequences (300 bits) under different pattern ratios for PDF data. (e.g., 0.4\% at the 4th row of [0, 0.1) indicates that 0.4\% of 300-bit sequences in the PDF dataset have the ratio of pattern `00' in its 300-bit sequence with 0\% - 10\%.)}
\begin{tabular}{|c||c|c|c|c|}
\hline
                                                          & '00'           & '01'           & '10'          & '11'          \\ \hline
Overall distribution                                                               & 25.7\%         & 24.6\%         & 24.6\%        & 25.1\%        \\ \hline \hline
\begin{tabular}[c]{@{}c@{}}Pattern ratio in \\ one sequence\end{tabular} & \multicolumn{4}{c|}{Ratio of sequences with the pattern ratios} \\ \hline
{[}0, 0.1)                                                               & 0.4\%          & 0.0\%          & 0.0\%         & 2.4\%         \\ \hline
{[}0.1, 0.2)                                                             & 20.1\%         & 6.0\%          & 6.0\%         & 17.7\%        \\ \hline
{[}0.2, 0.22)                                                            & 11.7\%         & 9.7\%          & 9.7\%         & 11.7\%        \\ \hline
{[}0.22, 0.24)                                                           & 13.6\%         & 24.6\%         & 24.6\%        & 13.6\%        \\ \hline
{[}0.24, 0.26)                                                           & 13.7\%         & 33.4\%         & 33.4\%        & 13.7\%        \\ \hline
{[}0.26, 0.28)                                                           & 12.0\%         & 19.3\%         & 19.3\%        & 11.9\%        \\ \hline
{[}0.28, 0.3)                                                            & 9.2\%          & 5.4\%          & 5.4\%         & 9.4\%         \\ \hline
{[}0.3, 0.4)                                                             & 15.5\%         & 1.5\%          & 1.5\%         & 17.5\%        \\ \hline
{[}0.4, 1{]}                                                             & 3.8\%          & 0.0\%          & 0.0\%         & 2.2\%         \\ \hline
\end{tabular}\label{tab:pdf}
\color{black}
\end{table}

\begin{figure}[!t]
	\centering
	\includegraphics[width=3.3in]{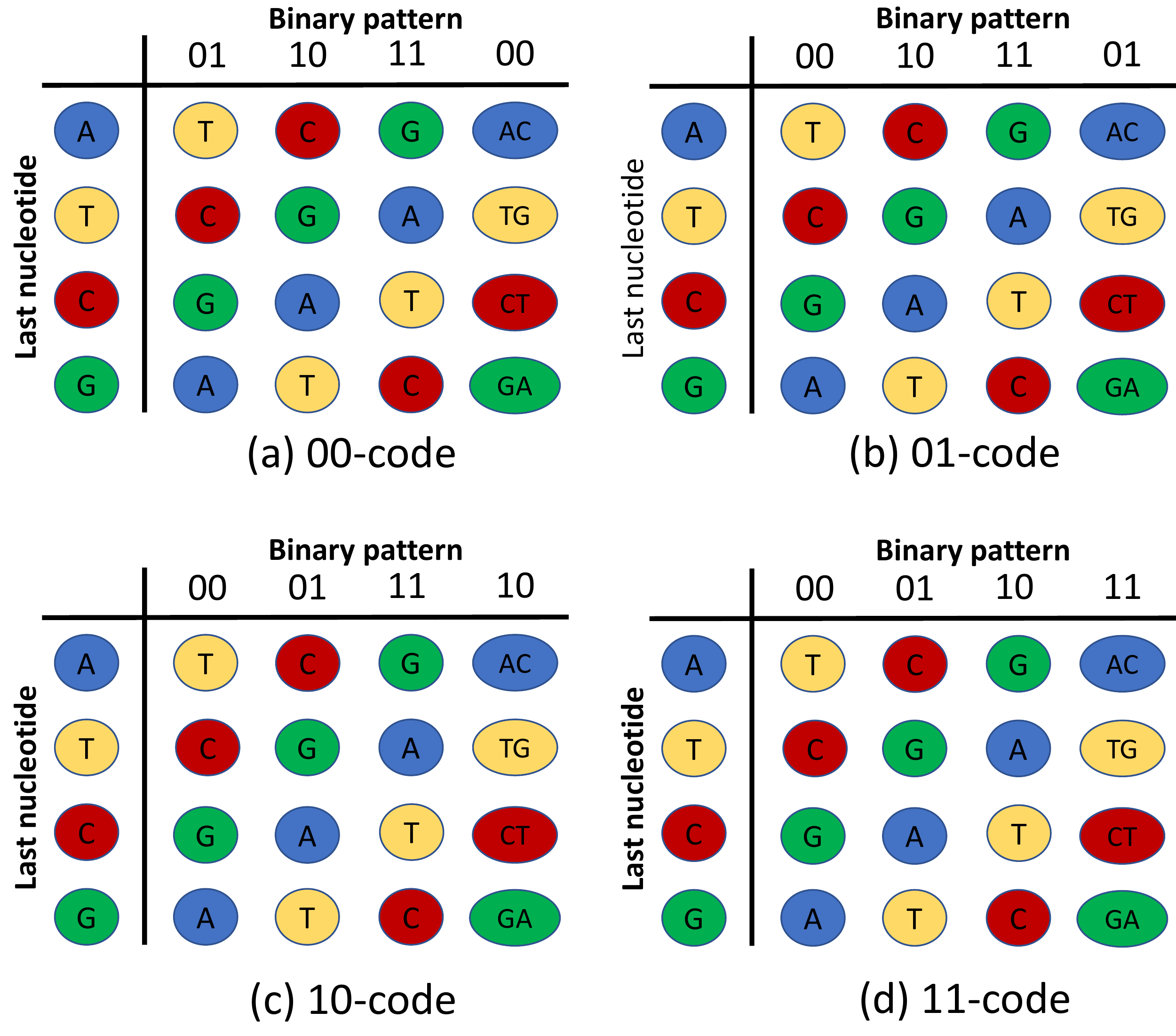}
	\caption{The extended unbalanced codes with 00-code, 01-code, 10-code and 11 code.}
	\label{fig:Extended_unbalance}
\end{figure}

As motivated by the non-uniform distributions of digital patterns in binary sequences, we apply different encoding schemes for binary sequences to further improve the encoding density of DNA storage. Since we have a binary sequence for each DNA strand before encoding, we know the frequencies of those 2-bit binary patterns in the binary sequence. So, we can encode those higher-frequency patterns into fewer nucleotides. To achieve that, we extend the 11-code in Figure~\ref{fig:coding_combined} (b) to generate a set of codes for different digital patterns (i.e., 00-code, 01-code, 10-code, and 11-code) as shown in Figure~\ref{fig:Extended_unbalance} called Digital Pattern-Aware Codes (DPAC). xx-code indicates that the column of the binary pattern `xx' has two nucleotides and other columns (patterns) have one nucleotide. Based on that, we can find that xx-code is not a good choice for the binary pattern `xx' but can achieve higher encoding density for other patterns. Therefore, according to the frequencies of binary patterns in a binary sequence, we first find the lowest-frequency binary patterns among those four binary patterns (i.e., '00', '10', '01' and '11'). Then, we will apply the corresponding code to this binary sequence. By doing so, it will minimize the effect of encoding two bits to two nucleotides as discussed previously. For example, for the first 300 bits of the image in Figure~\ref{fig:digit_distribution} (b), the pattern `01' has the lowest frequency, and thus we use 01-code for this first 300 bits. To analyze the worst case of the pattern-aware code method, if we have a uniform distribution for those four digital patterns in a sequence (i.e., 25\% ratio for each binary pattern), we will obtain the lowest encoding density which is 1.6 bits/nt. Otherwise, we will get a higher encoding density than 1.6 bits/nt. Therefore, even in the worst case, the lowest encoding density of DPAC scheme is still higher than the average encoding densities of prior work~\cite{bornholt2016dna, organick2018random}.  

\subsection{Adding 2bit-code \color{black}}\label{sec:2bit-code}
Since the 2bit-code in Figure~\ref{fig:coding_combined} has the encoding density of 2 bits/nt, we desire to use it enhancing the overall encoding density of DNA storage further. Previously, the reason that people cannot directly use 2bit-code is the code will potentially generate long homopolymers (e.g., more than four identical consecutive nucleotides). Especially for a large amount data, the possibility of having a long homopolymers in DNA sequences encoded by 2bit-code is high. These homopolymers will cause much high error rates and may result in wrong sequencing results. 

Due to the multi-encoding manner in the DPAC code, it becomes possible to use 2bit-code in the DP-DNA scheme. The principle of adding 2bit-code is that, by using 2bit-code, avoiding long homopolymers in each individual DNA sequence is much easier than that of purely using 2bit-code for all dataset, since DNA strand length is only around 100bp - 300bp in DNA storage. To avoid the homopolymers encoded from the 2bit-code, for each DNA strand, we can first investigate the feasibility of the 2bit-code and then check whether the 2bit-code generates any long homopolymers. If not, the binary sequence will be encoded by the 2bit-code. Otherwise, we discard the 2bit-code for this DNA strand, and use the DPAC code to encode it. By doing so, some DNA strands can be encoded with 2 bits/nt, and thus can further increase the overall encoding density of DNA storage.


\subsection{Variable-length (VL) Scheme}\label{sec:variablelength}
Based on the all above discussion, we assume that all DNA strands follow the same strand length. However, the DNA length also has the large influence on the encoding density from two perspectives. First, since one DNA strand contains a portion of metadata, the ratio between the metadata and the payload in DNA strands determines the overall encoding density. It means that the large payload ratio can obtain a higher overall encoding density. On the other hand, the strand length decides the cutting points in a long binary sequence. The cutting positions may have effect on the choice of the coding schemes.

Specifically, for the first perspective, normally longer DNA strand length results in a higher overall encoding density. But, due to the constraints of current synthesis and sequencing technology, longer DNA strand length may cause higher errors or even make sequencing failed. So, the most studies used about 100-300 bps as the DNA strand length. With the development of biochemical technology, the DNA strand length is expected to become longer. The investigate of the DNA strand length can be found in Section~\ref{sec:VL}. For the second perspective, prior work currently only used one specific encoding scheme and thus their schemes do not affect the payload encoding density under a fixed encoding scheme. However, by using multiple encoding schemes, the cutting points really do matter on which encoding scheme to be used. For example, for a binary sequence, the first 390 bits can be encoded into an encoding scheme with higher encoding density (e.g., 2bit-code), but the first 400 bits can only be encoded with a lower-density code since the following 10 bits (i.e., 391-400 bits) with the higher-density code will face the feasibility issue. In this scenario, we may want to cut the first 390 bits and encode it to a higher-density code but a shorter length. The rest 10 bits can be left for the following DNA strand. By doing so, the overall encoding density may be increased due to using a higher-density code.

Therefore, to further increase the overall encoding density of DNA storage, we propose a variable-length (VL) scheme. As discussed above, the purpose of variable length is to enable use a higher-density code although it may shorten the DNA strand length. There is a tradeoff between the DNA strand length and encoding schemes. Since one DNA strand contains metadata, which is a fixed length for a DNA storage system, a shorter DNA strand length causes a lower overall encoding density (i.e., the ratio of the length of binary sequence to the DNA strand length). To achieve a high overall encoding density, we use a constraint check to determine whether an early cut can make the DNA strand encoded by 2bit-code obey the constraints. Also, we need to check if the higher-density code can finally achieve a higher overall encoding density than that of a lower-density code as indicated in Equation~(\ref{eq:variable}).
\begin{equation}\label{eq:variable}
\frac{L}{L/\epsilon_1 + L_{meta}} < \frac{L-M}{(L-M)/\epsilon_2 + L_{meta}}
 \end{equation}
where $\epsilon_1$ and $\epsilon_2$ indicate the code densities of the low-density and high-density codes, respectively. $L$ is the default length of binary sequence to be encoded. $M$ is indicates how many bits are excluded for the high-density code. $L_{meta}$ refers to the number of nucleotides used for metadata such as primer pairs and internal index in DNA strands. Thus, based on the equation, if we can find the early cut can increase the encoding density of the new DNA strand, we will use the high-density code (i.e., 2bit-code). 

\subsection{DP-DNA Encoding}\label{sec:hybrid}
\begin{figure}[!t]
	\centering
	\includegraphics[width=3.3in]{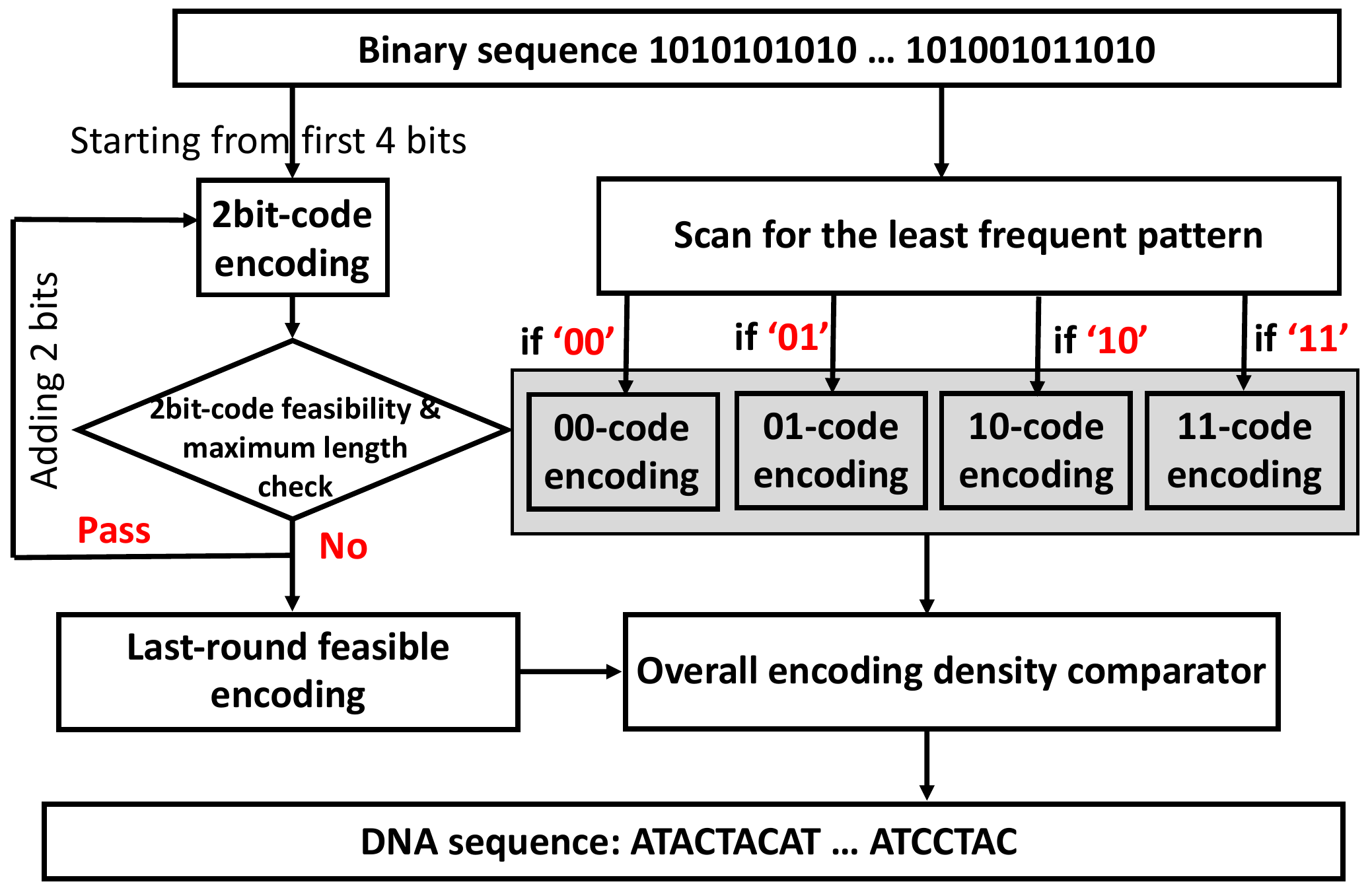}
	\caption{Design flow of the DP-DNA encoding scheme.}
	\label{fig:flow}
\end{figure}
Based on the previous discussion, we propose a hybrid encoding scheme for DNA storage systems. For the encoding process, as shown in Figure~\ref{fig:flow}, to encode a binary sequence, there are two paths. One is to perform the binary pattern to find the binary pattern with the lowest frequency, and then select the corresponding code as a candidate. For example, if the pattern `11' has the lowest frequency, 11-code will be selected. The other path is to find the maximum feasible length of 2bit-code by encoding the binary sequence from the first bit. Finally, we can find a proper length for the DNA strand with the 2bit-code. After those two paths, two encoding manners will be compared to check which one provides a higher overall encoding density as indicated in Equation~(\ref{eq:variable}), and the proper code will be finally used for this binary sequence. Note that $M$ (i.e., how many bits are excluded) is obtained from the 2bit-code feasibility checking process, $\epsilon_{2}=2bits/nt$, and $\epsilon_{1}$ is calculated from the first path.

Since we use five individual encoding schemes for the binary sequences, we propose a new DNA encoding format to decode DNA strands to distinguish their encoding algorithms. As shown in Figure~\ref{fig:DNA_architecture}, a new field `Encoding' is added in the DNA strand format. In the Encoding field, five encoding schemes can be represented by binary digits 0000, 0001, 0010, 0100, and 0101, respectively. We encode the Encoding field using 11-code. So, two nucleotides in the encoding field are enough to distinguish these five encoding schemes (missing 0011 here avoids needing one more nucleotide to be used since `11' needs two nucleotides in 11-code and thus 0011 will need three nucleotides). So, after finding the code as shown in Figure~\ref{fig:flow}, we use the corresponding encoding scheme to encode the binary sequence and the corresponding `Encoding' field is also recorded. For the decoding process, compared to the traditional decoding process, one extra step is needed. After sequencing, we first need to use 11-code to decode the first two nucleotides (after the primer) and to identify the encoding scheme of this DNA strand. Then, the rest of the DNA strand can be fully decoded by the corresponding coding scheme.

\subsection{Overall DP-DNA System}
\begin{figure}[!t]
	\centering
	\includegraphics[width=3.3in]{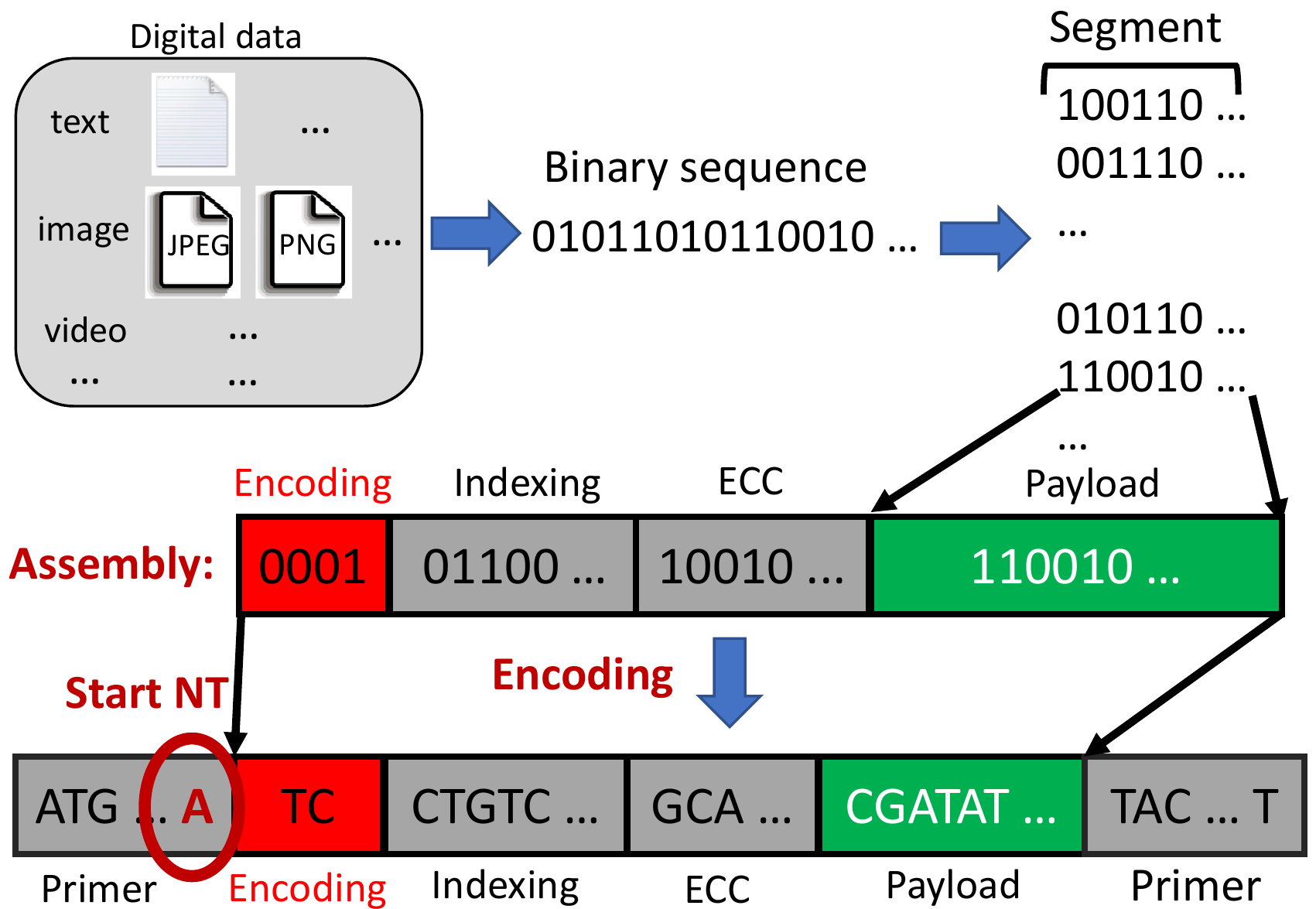}
	\caption{The newly proposed DNA strand encoding format.}
	\label{fig:DNA_architecture}
\end{figure}

In this subsection, we introduce the overall DP-DNA system. In the previous studies, they first encode a long digital sequence into a long DNA sequence. Then, the DNA sequence is chunked to short DNA sequences (payloads) with the same length~\cite{goldman2013towards, organick2018random, blawat2016forward, tomek2019driving}. After that, the other fields such as primers and an internal index are attached to the payloads. Therefore, DNA strands in their work are dependent on other DNA strands. If a nucleotide in a DNA strand (assume the DNA strand is in the middle of the long DNA sequence) is sequenced wrongly, other DNA strands following this strand will also be decoded wrongly.

In our proposed scheme, as shown in Figure~\ref{fig:DNA_architecture}, first we chunk a binary sequence into small segments (e.g., 300 bits). The segment size is related to the final DNA strand length. Based on current biotechnology, the DNA strand length is 100 bp - 300 bp. Thus, the segment size can be around 100 bits - 400 bits according to different encoding densities and DNA strand lengths. The segment size can be fixed at 300 bits or can be a variable size based on its contexts according to the requirement. After chunking a binary sequence into segments, those segments and other information, including primers, Encoding, ECC and internal index, are assembled. Finally, we encode these assembled binary sequences to DNA strands. Each DNA strand has a `Start NT' which is the last nucleotide of the primer. The `Start NT' nucleotide is used as the first nucleotide for the rotating encoding, and the following nucleotides will be encoded one by one based on its previous one. Thus, each DNA strand has its own `Start NT' and each DNA strand is independent to other DNA strands. There are two advantages. One is that the independent DNA strands can potentially speed up the encoding process since all DNA strands can be encoded in parallel. However, for the previous studies, since the DNA strands are encoded based on their previous DNA strand, it is hard to make the encoding process in parallel. The second is that the previous studies may face the scenario of error propagation, which means that one DNA nucleotide error may cause errors for its following DNA nucleotides since its following DNA nucleotides are encoded based on their previous nucleotides. As a result, although the following DNA nucleotides can be read out correctly, the binary sequence will be decoded wrongly if the nucleotide has an error. So, 'Start NT' has a great potential for those approximate DNA storage system to increase the robustness of DNA storage system. The `Encoding' field will be encoded by 11-code as discussed in Section~\ref{sec:hybrid}. After that, the following fields will be encoded based on the encoding scheme indicated by the `Encoding' field.


In summary, the proposed DP-DNA scheme has a three-fold optimization on the encoding density. One is that the unbalanced code itself (e.g., 11-code) increases an average density to 1.6 bits/nt (1.57 bits/nt in~\cite{goldman2013towards}). The second one is that the multi-DPAC code can use a code (i.e., 2bit-code) with the highest encoding density. It can further improve the overall encoding density of DNA storage. The feasibility check of using the 2bit-code helps avoid the violations of biochemical constraints. Finally, according to the non-uniform distribution of different two-bit patterns, we dynamically apply different encoding schemes to each binary sequence to increase the encoding density.

\section{Feasibility, Overhead, and Random Access Discussion}\label{sec:overhead}
The experiments of this work are based on simulation. The wet-lab validation experiment will be left for future work. Although the wet-lab experiments can validate the proposed schemes, the feasibility check is a common technique in biological fields such as primer design, and it can broadly reflect real-life events and improve the success rate of wet-lab experiments. Moreover, the cost of wet-lab experiments in DNA storage is extremely expensive (e.g., 1GB data may cost more than \$1,000,000~\cite{kosuri2014large}) and thus the wet-lab experiments lacks scalability to cover all possible cases especially for large datasets. The simulation can investigate all possibilities of different schemes and provide the scalability of DNA storage investigation. 

To ensure the feasibility of the proposed scheme, we design a set of feasibility rules to check synthesis and sequencing efficiency based on commercial design rules~\cite{DNA_rules, Synthesis_rule} and previous studies~\cite{bornholt2016dna, erlich2017dna, organick2018random}. The feasibility rules are shown in the following:
\begin{itemize}[leftmargin=10pt, topsep=1pt,itemsep = 0pt]
	\item Absence of long homopolymers (less than four nucleotides)~\cite{erlich2017dna, blawat2016forward}.
	\item GC contents (40\% - 60\%)~\cite{DNA_rules, organick2018random, addgene}.
	\item Avoiding Hairpin Structures: trys to avoid sequences containing two inverted repeats, separated by at least three nucleotides~\cite{DNA_rules}.
	\item DNA strand length smaller than 1000 bp~\cite{DNA_rules, Synthesis_rule}.
\end{itemize}
\begin{table}[!t]
	\small
	\centering
	\tabcolsep = 5pt
	\caption{Results of design rules check for DP-DNA and the reference work~\cite{organick2018random}.}
	\begin{tabular}{|c|c|c|}
		\hline
		& Organick et al.~\cite{organick2018random}    & DP-DNA \\ \hline
		GC content (Ave.)          & 50.01\% & 50.02\% \\ \hline
		long homopolymer violation & No      & No      \\ \hline
		Long DNA strand length     & No      & No      \\ \hline
		Hairpin structure ratio$^{*}$   & 0.0062     & 0.0065    \\ \hline
		Average score per strand               & 0.0050     & 0.0052     \\ \hline
	\end{tabular}\label{tab:design_rule}
	\begin{tablenotes}\footnotesize
		\item \footnotesize{$^{*}$Hairpin structure ratio refers to the number of DNA strands containing hairpin structures divided by the total number of DNA strands.}
	\end{tablenotes}
\end{table}
To check the feasibility of encoding schemes, we set a score by these rules to indicate whether an encoding scheme will have a good design for real synthesis and sequencing. Initially, the score is zero. Then, the encoded DNA strands will go through the rules, and any violations of each rule will make the score increased by one. For example, if a 300 bp DNA strand has 63\% GC content and contains a DNA pattern `AAAAA', this DNA strand violates the first two rules. So, the score of this DNA strand is 2. Therefore, the lower the scores are, the higher efficiency the synthesis (write) and sequencing (read) processes will have. In other words, lower scores can make the DNA write and read processes of DNA storage easier. Although these rules cannot 100\% ensure that DNA strands with the score of zero can be successfully synthesized or sequenced, the design rules are widely used in DNA techniques and highly related the success rate of synthesis and sequencing. Therefore, generally a smaller score indicates a higher success rate of synthesis and sequencing processes in wet-lab experiments.

We use the work in~\cite{organick2018random} as a reference, which implemented wet-lab experiments for their DNA encoding scheme. Before wet-lab experiments, they also used a similar set of rules to design their primers and check biochemical violations. So, we assume that if our proposed scheme achieves a similar or lower score than that of the work~\cite{organick2018random}, we can say that our proposed scheme will likely be successfully synthesized and sequenced according to current biology techniques. We use different data sets discussed in Section~\ref{sec:result} and split data into around 300 bp DNA strands (the reason we choose 300 bp is that a longer DNA strand length is more likely to violate those design rules). Then, we compute the score of each DNA strand. As shown in Table~\ref{tab:design_rule}, both schemes have around 50\% GC contents due to the distribution of 'G' and 'C' nucleotides in the coding schemes each being 25\%. Moreover, both schemes can avoid long homopolymers due to rotating encoding behavior. For the hairpin structure, we expected the number of hairpin structures in DNA strands as small as possible. As shown in Table~\ref{tab:design_rule}, both schemes have similar hairpin structure ratios, and thus have similar sequencing difficulty in that regard. As indicated in Table~\ref{tab:design_rule} the proposed DP-DNA scheme achieves a slightly higher score per DNA strand than the referenced work~\cite{organick2018random}. The results indicate that those two schemes have much similar difficulty to read original DNA strands out. 

For the overhead discussion of this scheme, the main overhead is from the computation. Our experiments are based on Python 3.6.9 running in a system with Intel(R) Xeon(R) CPU E5-2620 with 2.4GHz and 32GB memory. Based on our experiments, to encode a DNA strand with 150 bp length, on average, the DP-DNA needs 91.4us and the baseline~\cite{organick2018random} needs 82.4us, which is about 10.9\% faster than that of DP-DNA. The reason is that the DP-DNA needs to check the 2-bit-code feasibility and scan the digit patterns, but those operations are much simple. However, the computation time of DP-DNA can be pipelined with DNA synthesis time. More importantly, compared to current DNA synthesis time (e.g., about 10s hours), the encoding overhead is acceptable. Moreover, by changing to a more efficient computing manner (e.g., running in C/C++, running in parallel, and running in high-performance systems), the execution time of the proposed encoding scheme will be tremendously decreased and can probably be ignored compared to the required DNA synthesis time.

For the random access of this design, we directly use the primer design rules in~\cite{organick2018random} to generate the primer pool for the DNA storage. Also, the architecture of this design maintains primer pairs and internal index in DNA strands, which is similar to the work~\cite{organick2018random}. Therefore, when randomly accessing a specific file, a mapping table can tell us the primer pairs of the DNA strands containing the target files. Then, Polymerase Chain Reaction (PCR) will amplify all DNA strands with the primer pairs. After that, those DNA strands will be read out by the sequencing machine. Finally, according to the internal index, the target files will be decoded and read out.\color{black}
\section{Experimental Results}\label{sec:result}
\subsection{Environment Description}
This section introduces the experimental results to explore the encoding density of the proposed scheme. The default DNA strand length is 150 bp (i.e., the maximum DNA strand length) for all schemes. Note that for DP-DNA the DNA strand lengths may be changed since the encoding densities of DNA strands are varied based on the distribution of different patterns in binary sequences. We use 20 bp for the length of the beginning and ending primers in a DNA strand. The ECC has 15\% overhead and we assume the ECC can make sure the data correctly read out as indicated in~\cite{organick2018random}. We use around 52GB data including images~\cite{imagenet} (about 3.4GB), file documents including text files, MSR storage trace files (csv type)~\cite{narayanan2008write} and IEEE/ACM PDF papers (about 5.8GB), video from YouTube (about 22GB), TPC-H database~\cite{TPCH} (20GB) and web pages (about 1GB) for the experiments. \color{black}

To investigate the encoding density, we use three baselines denoted by Church et al.~\cite{church2012next}, Organick et al.~\cite{organick2018random} and Blawat et al.~\cite{blawat2016forward}, respectively. For the proposed DP-DNA, two options are investigated with different maximum numbers (X) of consecutive identical nucleotides (homopolymers) denoted by Homo-X. All these five schemes are implemented in the experiments and applied to all data including video, text, web, database and image. Two metrics, payload encoding density and overall encoding density (including internal index, ECC and primers), are used based on these encoding schemes.

\begin{figure}[!t]
	\centering
	\includegraphics[width=3.3in]{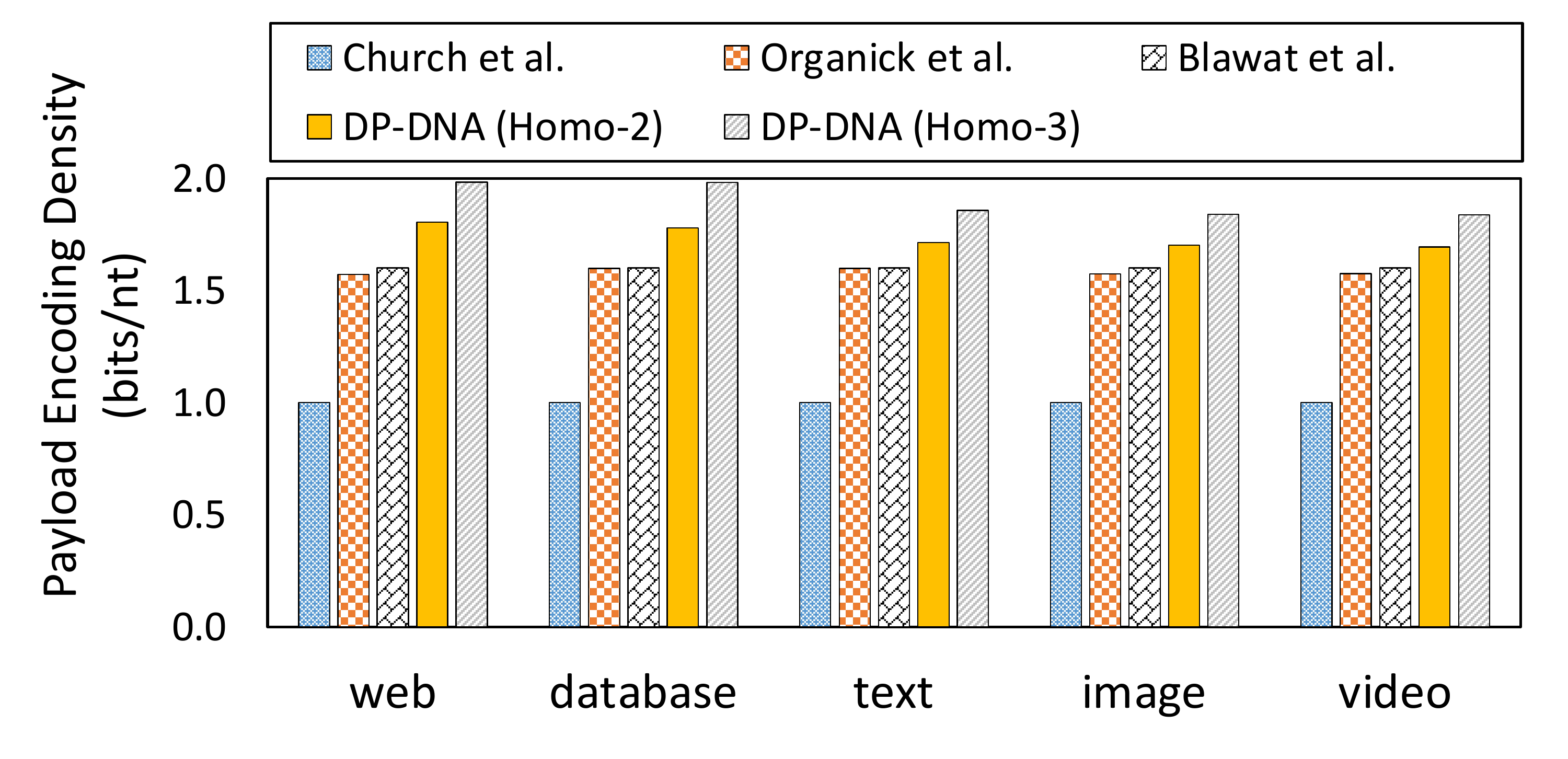}
	\caption{The comparison of payload encoding densities for different schemes.}
	\label{fig:Payload_Encoding}
\end{figure}

\begin{figure}[!t]
	\centering
	\includegraphics[width=3.3in]{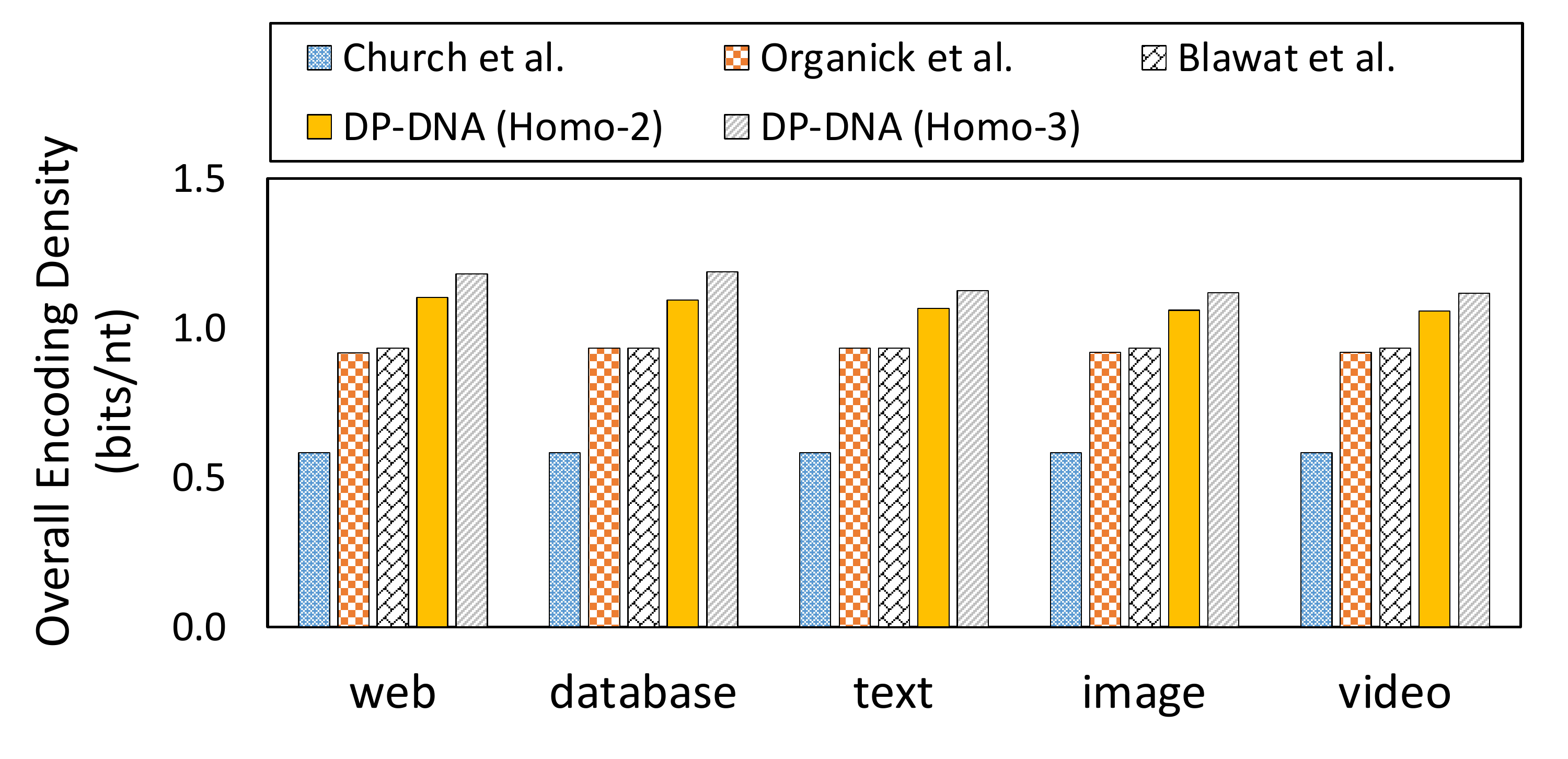}
	\caption{Overall encoding densities with different schemes.}\label{fig:Overall_Encoding}
\end{figure}

\subsection{DNA Storage Overall Encoding Density}
First, we investigate the payload encoding density. As shown in Figure~\ref{fig:Payload_Encoding}, we can find that the proposed DP-DNA (Homo-2) and DP-DNA (Homo-3) can achieve 5.9\% - 80.3\% and 14.9\% - 98.4\% encoding density improvement compared to the three baselines, respectively. Homo-3 has a higher encoding density than that of Homo-2 because Homo-3 obtains more passes on the 2bit-code's feasibility check than Homo-2. Compared with the three baselines, there are two main reasons for DP-DNA to achieve higher encoding density. The first reason is that the digital pattern-aware coding (DPAC) based on the distribution of two-bit patterns in a binary sequence can distinguish low-frequency and high-frequency patterns. Those high-frequency patterns are encoded with 2 bits/nt (the upper bound storage density) and only the least frequent pattern is encoded with 1 bit/nt. Another reason is that the hybrid encoding manner provides the possibility of using 2-bit code for some DNA strands which further enhance the overall encoding density since the 2-bit code provides the highest encoding density. Moreover, the variable-length scheme increases the possibility of using 2bit-code in DNA strands. Thus, compared to the existing work~\cite{organick2018random, church2012next, blawat2016forward}, DP-DNA can significantly improve the encoding density with a range of 1.694 bits/nt - 1.984 bits/nt.

As shown in Figure~\ref{fig:Overall_Encoding}, the overall encoding density is the encoding density including the primer pair, Encoding, ECC, payload, and internal index. Thus, the overall encoding densities are smaller than the payload encoding densities. Compared to the previous studies, although the DP-DNA scheme adds two extra nucleotides for the Encoding field, the proposed DP-DNA scheme can still achieve 13.1\% - 103.4\% encoding density improvement. 

\color{black}

\begin{figure}[!t]
	\centering
	\includegraphics[width=3.3in]{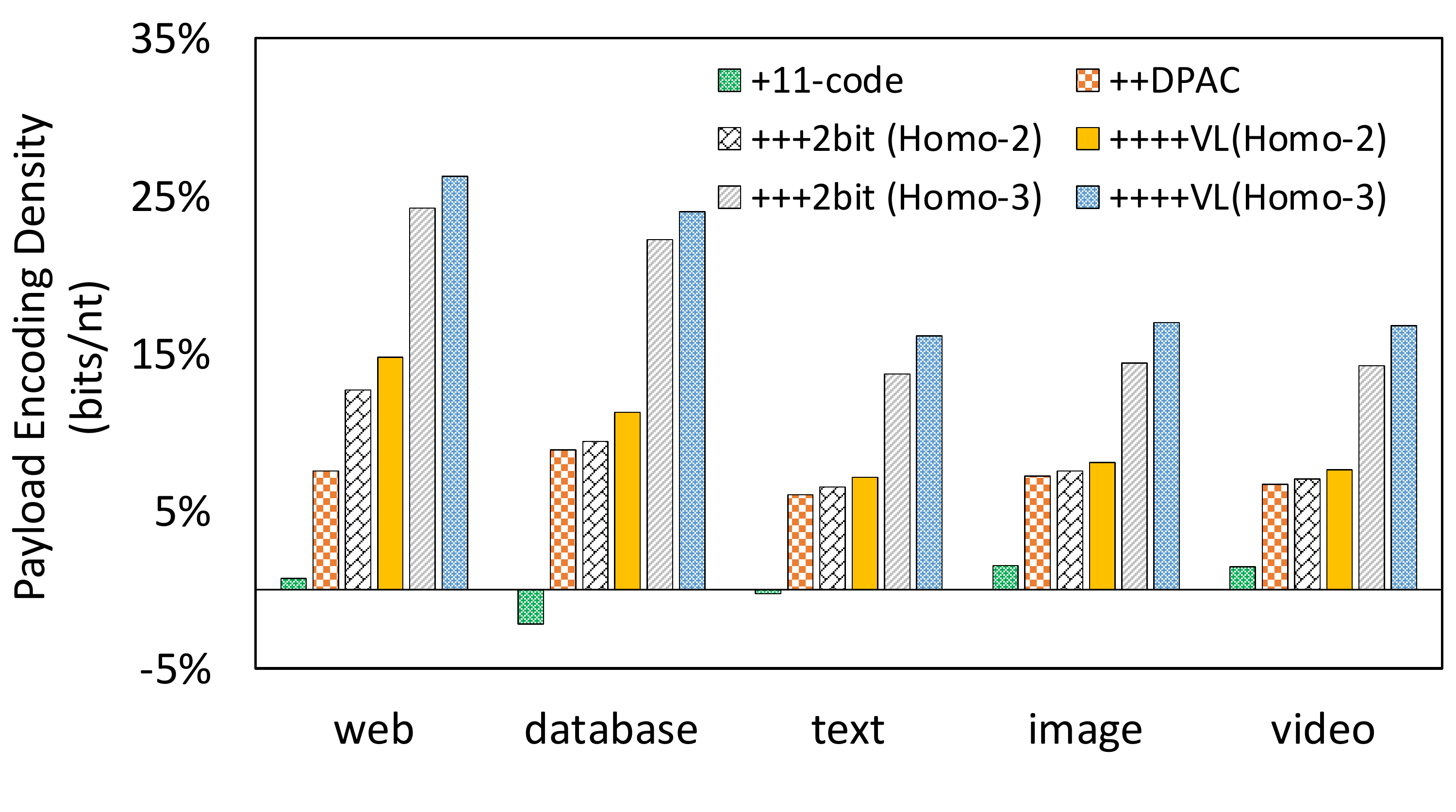}
	\caption{Breakdown analysis of the payload encoding density improvement for different applications compared to Organick's work~\cite{organick2018random}.}
	\label{fig:breakdown}
\end{figure}
\subsection{Breakdown Analysis}
In this subsection, we discuss the effect of different techniques on the payload encoding density. Starting from Organick's work~\cite{organick2018random} as a baseline, we add one feature each time until it becomes DP-DNA as described below:
\begin{itemize}[leftmargin=*]
	\item \textbf{Organick et al.~\cite{organick2018random}:} is the baseline scheme.
	\item \textbf{+11-code:} only uses 11-code for encoding binary sequences to DNA strands.
	\item \textbf{++DPAC:} considers binary patterns and uses the corresponding codes for those binary sequences.
	\item \textbf{+++2bit (Homo-X):} adds 2bit-code and uses 2bit-code feasibility check with the maximum of X consecutive identical nucleotides.
	\item \textbf{++++VL (Homo-X):} adds the variable length (VL) scheme (i.e., DP-DNA (Homo-2) and (Homo-3)).
\end{itemize}

We use different applications (i.e., web, database, text, images, and video) to compute the payload encoding density improvements for different schemes as shown in Figure~\ref{fig:breakdown}. We can find that compared to Organick's work, the density improvement of 11-code varies from -2.18\% to 1.54\% (i.e., 1.564 bits/nt (min) and 1.598 bits/nt (max)). The reason of causing such a variance from negative to positive improvement is that the `11' pattern may or may not be the pattern with the lowest ratio in DNA strands. If the `11' pattern has the lowest ratio, 11-code will obtain the highest encoding density since a small number of digits in bit streams are encoded into two nucleotides. In contrast, the encoding density will be decreased when encoding `11' pattern into two nucleotides with a high ratio. For DPAC, the encoding density improvement can be increase to 6.02\% - 8.87\%. This is because DP-DNA find the lowest-frequent pattern and encode them to two nucleotide. For other digital patterns, DP-DNA can use the highest encoding rate (i.e., two bits to one nucleotide). Finally, the variable-length scheme (++++VL) achieves significant improvement with the increase of 7.03\% - 26.23\%. The results benefit from two perspectives. One is the 2bit-code achieves the highest encoding density. The other is that the variable-length scheme increases the number of DNA strands encoded by 2bit-code. 
\color{black}

\subsection{Varying DNA Strand Length}\label{sec:VL}
\begin{figure}[!t]
	\centering
	\includegraphics[width=3.3in]{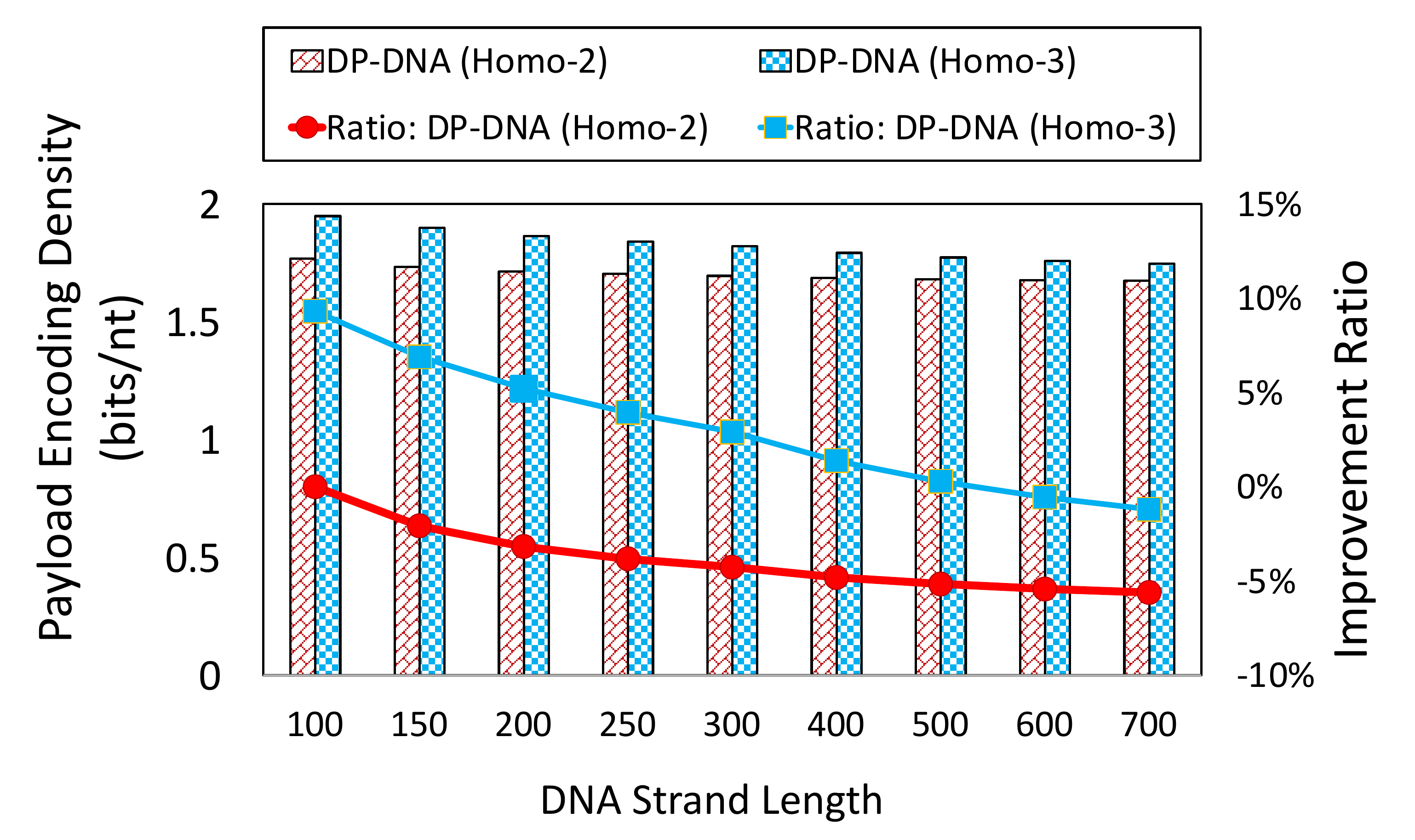}
	\caption{Payload encoding density and encoding improvement with varying DNA strand lengths. The encoding improvement is based on the payload encoding density of DP-DNA (Homo-2) with the DNA length 100nt for comparison.}
	\label{fig:coding_length}
\end{figure}
\begin{figure}[!t]
	\centering
	\includegraphics[width=3.3in]{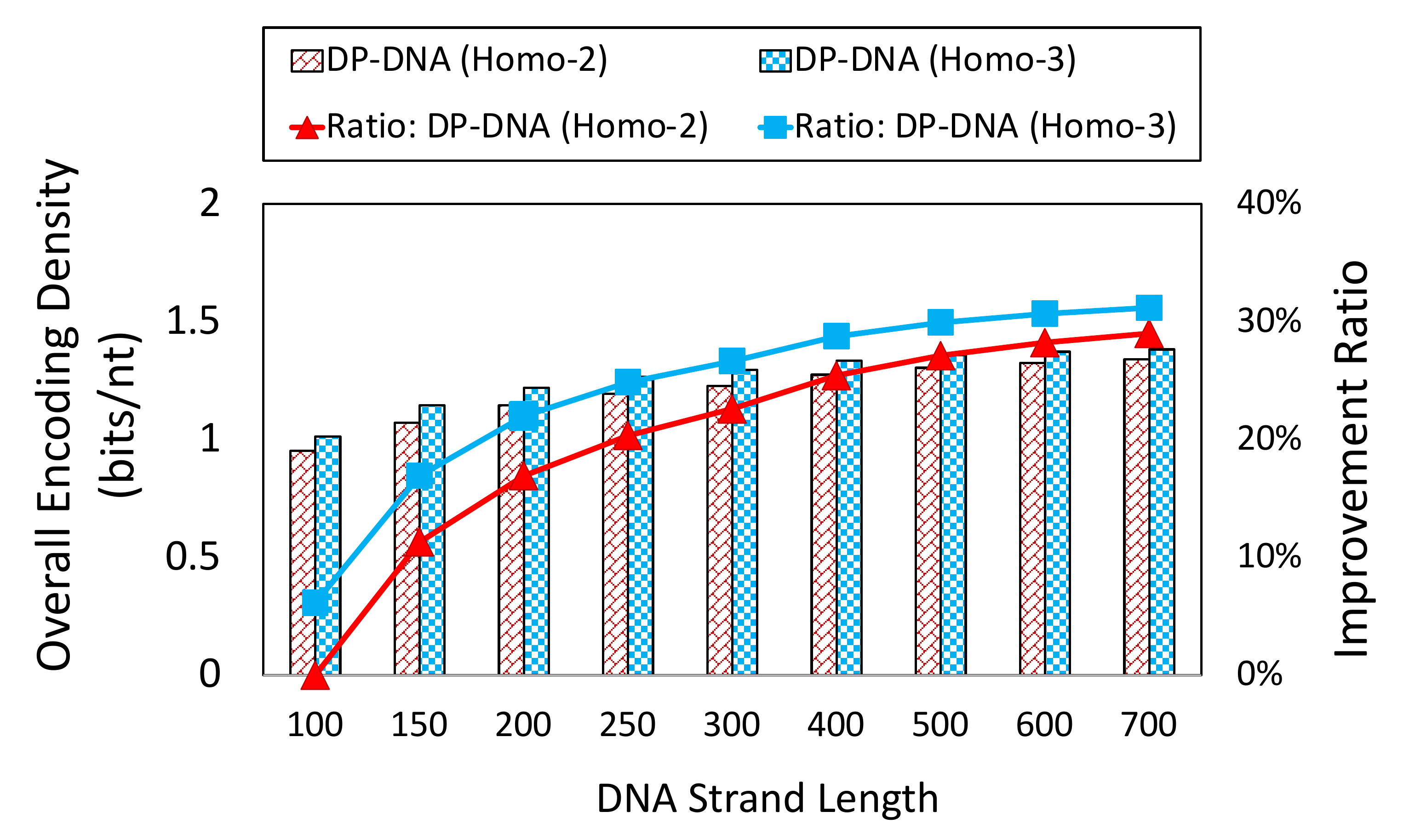}
	\caption{Overall encoding density with varying DNA strand lengths for image data. The encoding improvement is based on the payload encoding density of DP-DNA (Homo-2) with the DNA length 100nt for comparison.}
	\label{fig:coding_length_overall}
\end{figure}
In this subsection, we investigate the effect of DNA strand length on the encoding density for different schemes. We compute the DNA encoding density based on the whole dataset (i.e., around 52GB data) and the conclusions will remain similar for each individual data type. The payload encoding density with varying DNA strand length is shown in Figure~\ref{fig:coding_length}. Note that since DP-DNA uses DPCA and VL schemes which can cause variable length, the DNA strand length here indicates the maximum DNA strand length can be used. The absolute encoding density and improve ratio are used. For the improvement ratio, the encoding density of DP-DNA with 100nt DNA length as a reference for comparison. 


For the payload encoding density, as increasing the DNA strand length from 100 bp to 700 bp, the payload densities have a decreasing trend (from 1.77 bits/nt to 1.67 bits/nt and from 1.95 bits/nt to 1.75 bits/nt for DP-DNA (Homo-2) and DP-DNA (Homo-3), respectively). There are two reasons for this. One reason is that a shorter DNA sequence has a more uneven distribution of digit patterns (i.e., `00', `01', `10' and `11'). Our proposed DP-DNA based on the pattern aware code can have more benefits from those uneven distribution patterns. The other is that the longer DNA strand has a higher possibility of holding consecutive identical nucleotides when using the 2bit-code. Thus, the 2bit-code becomes harder to pass the feasibility check for a longer DNA sequence. 

For the overall encoding density, since the DNA strand contains metadata including primer pairs, ECC, internal index and Encoding, the overall density is smaller than the payload only density. However, as increasing the DNA strand length, the overall densities have an increasing trend (from 0.95 bits/nt to 1.34 bits/nt and from 1.01 bits/nt to 1.38 bits/nt for DP-DNA (Homo-2) and DP-DNA (Homo-3), respectively). The reason is that as increasing DNA strand length the ratio of payload in a DNA strand is proportionally increased. Those influence of density degradation caused by metadata is significantly decreased. For example, in a 100-bp DNA strand, the primer pairs (40 bp) occupy 40\% of DNA length resulting in a low overall encoding density. When the DNA length is increased to 700 bp, the overhead of primer pairs is only 5.7\%. Therefore, although the payload encoding scheme may face a slight encoding density decrease, the overall density density will be increased due to the small ratio of metadata. In summary, with the development of biological technology, the DNA strand length will be continuously increased and the bio-constraints might be loosen. According to that, our scheme can still keep the advantages of high encoding density. 

\section{Related Work}\label{sec:related}
DNA has become an attractive storage medium due to its high spatial density and long durability. Previous DNA storage studies mostly focused on the feasibility of storing binary data in DNA~\cite{luby2002lt, mackay2005fountain, wang2019high, gibson2010creation, grass2015robust, heckel2017fundamental, organick2018random, church2012next, blawat2016forward, bornholt2016dna, choi2018addition, bancroft2001long, yazdi2017portable}. They proposed different encoding and decoding schemes and did real wet-lab experiments to store digital data into DNA storage. For example, for an early study, Bancroft et al.~\cite{bancroft2001long} used two DNA classes to store different types of data. They encoded English characters and a space to three nucleotides and finally achieved an information density of 1.06 nucleotides per bit.
Church et al.~\cite{church2012next} proposed to encode `A' or `C' to 1 and `T' or `G' to 0 which can achieve 1 bit/nt. Some studies~\cite{goldman2013towards, organick2018random, blawat2016forward, tomek2019driving} used a Huffman code to encode base-3 digits and then encoded these Huffman encoded values to DNA sequences in a rotating manner. Blawat~et al.~\cite{blawat2016forward} encoded one-byte data into one or two nucleotides. With the constraints of the maximum number of consecutive identical nucleotides as 3, they finally can achieve 1.6 bits/nt. Yazdi et al.~\cite{yazdi2017portable} used random numbers to shuffle digital data to increase the density and robustness of DNA storage, but the metadata overhead introduced by random numbers limits the scalability of their scheme. Li et al.~\cite{li2021img} proposed an approximate DNA storage system for images to improve the robustness of DNA storage.
\color{black} 
However, all those encoding schemes are based on a static mapping scheme from digital to DNA layers. With a dynamic changed incoming digital data, their schemes may not be able to achieve the best encoding density due to the unawareness of digital patterns.


\section{Conclusion}\label{sec:conclusion}
In this paper, we propose a new digital pattern-aware DNA storage encoding scheme, called DP-DNA, which can store digital data in DNA more efficiently in terms of the DNA storage encoding density. DP-DNA maintains a set of codes. A digital pattern-aware code (DPAC) analyzes a binary sequence's two-bit patterns and selects a proper code to achieve a high encoding density. A new multi-encoding field added to the DNA encoding format can distinguish the encoding schemes used for those DNA strands and allow us to correctly decode DNA data back to the binary data. Moreover, a variable-length scheme is newly proposed to increase the possibility of using 2bit-code in DNA storage system, further increasing the encoding density. Experimental results based on binary data indicate that the proposed DP-DNA achieves around 5.9\% - 103.5\% higher encoding density of DNA storage than other existing schemes.


\bibliographystyle{plain}
\bibliography{refs}

\end{document}